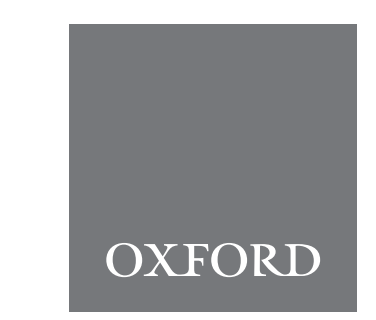

Sequence analysis

# GateKeeper: a new hardware architecture for accelerating pre-alignment in DNA short read mapping

**Mohammed Alser[1,*], Hasan Hassan[2,3], Hongyi Xin[4], Oğuz Ergin[2], Onur Mutlu[3,*] and Can Alkan[1,*]**

[1]Department of Computer Engineering, Bilkent University, Bilkent, Ankara 06800, Turkey, [2]TOBB University of Economics & Technology, Sogutozu, Ankara, Turkey, [3]Department of Computer Science, ETH Zürich, 8092 Zürich, Switzerland and [4]Department of Computer Science, Carnegie Mellon University, Pittsburgh, PA 15213, USA

*To whom correspondence should be addressed.

## Abstract

**Motivation:** High throughput DNA sequencing (HTS) technologies generate an excessive number of small DNA segments -called *short reads*- that cause significant computational burden. To analyze the entire genome, each of the billions of short reads must be mapped to a reference genome based on the similarity between a read and 'candidate' locations in that reference genome. The similarity measurement, called *alignment*, formulated as an approximate string matching problem, is the computational bottleneck because: (i) it is implemented using quadratic-time dynamic programming algorithms and (ii) the majority of candidate locations in the reference genome do not align with a given read due to high dissimilarity. Calculating the alignment of such incorrect candidate locations consumes *an overwhelming majority* of a modern read mapper's execution time. Therefore, it is crucial to develop a fast and effective filter that can detect incorrect candidate locations and eliminate them before invoking computationally costly alignment algorithms.

**Results:** We propose GateKeeper, a new hardware accelerator that functions as a *pre-alignment* step that quickly filters out most incorrect candidate locations. GateKeeper is the first design to accelerate pre-alignment using Field-Programmable Gate Arrays (FPGAs), which can perform pre-alignment much faster than software. When implemented on a single FPGA chip, GateKeeper maintains high accuracy (on average >96%) while providing, on average, 90-fold and 130-fold speedup over the state-of-the-art software pre-alignment techniques, Adjacency Filter and Shifted Hamming Distance (SHD), respectively. The addition of GateKeeper as a pre-alignment step can reduce the verification time of the mrFAST mapper by a factor of 10.

**Availability and implementation:** https://github.com/BilkentCompGen/GateKeeper

**Contact:** mohammedalser@bilkent.edu.tr or onur.mutlu@inf.ethz.ch or calkan@cs.bilkent.edu.tr

**Supplementary information:** Supplementary data are available at *Bioinformatics* online.

## 1 Introduction

High throughput sequencing (HTS) technologies are capable of generating a tremendous amount of sequencing data. For example, the Illumina HiSeq4000 platform can generate more than 1.5 trillion base pairs (bp) in less than four days. This flood of sequenced data continues to overwhelm the processing capacity of existing algorithms and hardware (Canzar and Salzberg, 2015). The success of the medical and genetic applications of HTS technologies relies on the existence of sufficient computational resources, which can quickly analyze the overwhelming amounts of data that the sequencers generate. An HTS instrument produces short reads (typically 75–150 bp) sampled randomly from DNA. In the presence of a reference genome, the short reads are first mapped to the long reference sequence. During this process, called *read mapping*, each short read is mapped onto one or more possible locations in the reference genome based on the similarity between the short read and the reference sequence segment at that location. Optimal *alignment* of the read and the reference segment could be calculated using the Smith-Waterman local alignment algorithm (Smith and Waterman, 1981). However, this approach is infeasible as it requires $O(mn)$ running time, where $m$ is the read length (100–150 bp for Illumina) and $n$ is the reference length ($\sim$3.2 billion bp for human genome), for each *read* in the dataset (hundreds of millions to billions). Therefore, read mapping algorithms apply heuristics to first find candidate map locations (*seed locations*) of subsequences of the reads using hash tables (Alkan *et al.*, 2009; David *et al.*, 2011; Hach *et al.*, 2010; Homer *et al.*, 2009; Xin *et al.*, 2013) or BWT-FM indices (Langmead and Salzberg, 2012; Langmead *et al.*, 2009; Li and Durbin, 2009; Li *et al.*, 2004), and then align the read in full *only* to those seed locations. Although the strategies for finding seed locations vary among different read mapping algorithms, seed location identification is typically followed by a *verification* step, which compares the read to the reference segment at the seed location to check if the read aligns to that location in the genome with fewer differences than a threshold. The verification step is the dominant part of the whole execution time in current mappers (over 90% of the running time) (Cheng *et al.*, 2015; Xin *et al.*, 2013). It calculates *edit distance* using quadratic-time algorithms such as Levenshtein's edit distance (Levenshtein, 1966), Smith-Waterman (Smith and Waterman, 1981) and Needleman-Wunsch (Needleman and Wunsch, 1970). Edit distance is defined as the minimum number of edits (i.e. insertions, deletions, or substitutions) needed to make the read exactly match the reference segment (Levenshtein, 1966). If the edit distance score is greater than a user-defined *edit distance threshold* (usually less than 5% of the read length (Ahmadi *et al.*, 2012; Hatem *et al.*, 2013; Xin *et al.*, 2015)), then the mapping is considered to be invalid (i.e. the read does not match the segment at seed location) and thus is rejected.

DEFINITION 1. *Given a candidate read r, a reference segment f, and an edit distance threshold E, the pairwise alignment problem is to identify a set of matches of r in f, where the read aligns with an edit distance* $\leq E$.

Recent work found that an overwhelming majority (>98%) of the seed locations exhibit more edits than the threshold (Xin *et al.*, 2013, 2015). These particular seed locations impose a large computational burden as they waste 90% of the mapper's execution time in verifying these incorrect mappings (Cheng *et al.*, 2015; Xin *et al.*, 2013). To tackle these challenges and bridge the widening gap between the execution time of the mappers and the increasing amount of sequencing data, most existing works fall into two approaches: (i) Design hardware accelerators to *accelerate the verification step* (Arram *et al.*, 2013; Houtgast *et al.*, 2015; Liu *et al.*, 2012; Luo *et al.*, 2013; Olson *et al.*, 2012; Waidyasooriya *et al.*, 2014). (ii) Build software-based *alignment filters* before the verification step (Cheng *et al.*, 2015; Marco-Sola *et al.*, 2012; Rasmussen *et al.*, 2006; Ukkonen, 1992; Weese *et al.*, 2009, 2012; Xin *et al.*, 2013, 2015). Such filters aim to minimize the number of candidate locations on which alignment is performed. They calculate a best guess estimate for the alignment score between a read and a seed location on the reference. If the lower bound exceeds a certain number of edits, indicating that the read and the segment at the seed location do not align, the seed location is eliminated such that no alignment is performed. Unfortunately, existing filtering techniques are either slow, such as Shifted Hamming distance (SHD) (Xin *et al.*, 2015), or inaccurate in filtering, such as the Adjacency Filter (Xin *et al.*, 2013) (implemented as part of FastHASH (Xin *et al.*, 2013)) and mrsFAST-Ultra (Hach *et al.*, 2014)). While mrsFAST-Ultra is able to detect only substitutions, FastHASH is unable to tolerate substitutions efficiently. We provide full descriptions of the key principles underlying each strategy in Supplementary Material, Section S1.2.

Our goal, in this work, is to minimize the mapper time spent on accurate alignment filtering. To this end, we introduce a new FPGA-based fast alignment filtering technique (called GateKeeper) that acts as a pre-alignment step in read mapping. To our knowledge, this is the first work that provides a new pre-alignment algorithm and architecture using reconfigurable hardware platforms. A fast filter designed on a specialized hardware platform can drastically expedite alignment by *reducing the number of locations that must be verified via dynamic programming*. This eliminates many unnecessary expensive computations, thereby greatly improving overall run time.

Our filtering technique improves and accelerates the state-of-the-art SHD filtering algorithm (Xin *et al.*, 2015) using new mechanisms and FPGAs. We build upon the SHD algorithm as it is the fastest and the most accurate filter (Xin *et al.*, 2015). Our new filtering algorithm has two properties that make it suitable for an FPGA-based implementation: (i) it is highly parallel, (ii) it heavily relies on bitwise operations such as shift, XOR and AND. Due to the highly parallel and bitwise-processing-friendly architecture of modern FPGAs, our design achieves more than two orders of magnitude speedup compared to the best prior software-based filtering approaches (SHD and Adjacency Filter), as our comprehensive evaluation shows (Section 3). Our architecture discards the incorrect mappings from the candidate mapping pool in a streaming fashion – data is processed as it is transferred from the host system. Filtering the mappings in a streaming fashion gives the ability to integrate our filter with any mapper that performs alignment, such as Bowtie2 (Langmead and Salzberg, 2012) and BWA-MEM (Li, 2013).

**Contributions.** We make the following contributions:

- We introduce the **first** hardware acceleration system for alignment filtering, called GateKeeper, which greatly reduces the need for alignment verification in DNA read mapping. To this end, we develop both a hardware-acceleration-friendly filtering algorithm and a highly parallel hardware accelerator design. We show that developing a hardware-based alignment filtering algorithm and architecture together is both feasible and effective by building our accelerator on a modern FPGA system.
- We comprehensively evaluate GateKeeper and compare it to two state-of-the-art software-based alignment filtering algorithms. A key result is that our design for reads of length 100 bp on a single FPGA chip provides, on average, 90-fold and 130-fold speedup over the state-of-the-art filters, Adjacency Filter (Xin *et al.*, 2013) and SHD (Xin *et al.*, 2015), respectively. Experimental results on both simulated and real datasets demonstrate that GateKeeper has a low false positive rate (the rate of incorrect mappings that are accepted by the filter) of 4% on average.
- We provide the design and implementation of a complete FPGA system and release its source code. To our knowledge, GateKeeper is the first open-source, freely available FPGA based alignment filter for genome analysis.

## 2 Gatekeeper architecture

### 2.1 Overview of our accelerator architecture

Based on the discussion provided in the Supplementary Material, Section 1.2, we introduce the **first** specialized FPGA-friendly hardware architecture for a *new alignment filtering algorithm*. The overall architecture, implementation details and flowchart representation of GateKeeper are discussed in the Supplementary Material, Section 1.3.1. Our current filter implementation relies on several optimization methods to create a robust and efficient filtering approach. At both the design and implementation stages, we satisfy several requirements: (i) Ensuring a lossless filtering algorithm by preserving all correct mappings. (ii) Supporting both Hamming distance and edit distance. The Hamming distance is a special case of the edit-distance. It is defined as the minimum number of substitutions required to change the read into the reference segment. The Hamming distance is computed in linear time. (iii) Examining the alignment between a read and a reference segment in a fast and efficient way (in terms of execution time and required resources).

### 2.2 Parallelization

GateKeeper is designed to utilize the large amounts of parallelism offered by FPGA architectures (Aluru and Jammula, 2014; Herbordt *et al.*, 2007; Trimberger, 2015). The use of FPGAs can yield significant performance improvements, especially for massively parallel algorithms. FPGAs are the most commonly used form of reconfigurable hardware engines today, and their computational capabilities are greatly increasing every generation due to increased number of transistors on the FPGA chip. An FPGA chip can be programmed (i.e. configured) to include a very large number of hardware execution units that are custom-tailored to the problem at hand. We take advantage of the fact that alignment filtering of one read is *inherently independent* of filtering of another read. We therefore can examine many reads in a parallel fashion. In particular, instead of handling each read in a sequential manner, as CPU-based filters (e.g. SHD) do, we can process a large number of reads at the same time by integrating as many hardware filtering processing cores as possible (constrained by chip area) in the FPGA chip. Each processing core is a complete alignment filter and can handle a single read at a time. We use the term 'processing core' in this paper to refer to the entire operation of the filtering process involved in GateKeeper. Processing cores are part of our architecture and are unrelated to the term 'CPU cores' or 'threads'.

### 2.3 GateKeeper processing core

Our primary purpose is to enhance the state-of-the-art SHD alignment filter such that we can greatly accelerate pre-alignment by taking advantage of the capabilities and parallelism of FPGAs. To achieve our goal, we design an algorithm inspired by SHD to reduce both the utilized resources and the execution time. These optimizations enable us to integrate more processing cores within the FPGA chip and hence examine many alignments at the same time. We present three new methods that we use in each GateKeeper processing core to improve execution time. Our first method introduces a new algorithmic method for performing alignment very rapidly compared to the original SHD. This method provides: (1) fast detection for exact matching alignment and (2) handling of one or more base-substitutions. Our second method supports calculating the edit distance with a new, very efficient hardware design. Our third method addresses the problem of hardware resource overheads introduced due to the use of FPGA as an acceleration platform. We provide the workflow of GateKeeper including the three optimization methods in the Supplementary Material, Figure S8. All features are implemented within the filtering processing core hardware and thus are performed highly efficiently. Next, we describe the three new methods.

#### 2.3.1 Method 1: Fast approximate string matching

We first discuss how to examine the alignment of reads against the reference sequence with a given Hamming distance threshold, and later extend our solution to support edit distance. Our first method aims to quickly detect the obviously-correct alignments that contain no edits or only few substitutions (i.e. less than the user-defined threshold). If the first method detects a correct alignment, then we can skip the other two methods but we still need the optimal alignment algorithms. A read is mappable if the Hamming distance between the read and its seed location does not exceed the given Hamming distance threshold. Hence, the first step is to identify all bp matches by calculating what we call a Hamming mask. The Hamming mask is a bit-vector of '0's and '1's representing the comparison of the read and the reference, where a '0' represents a bp match and a '1' represents a bp mismatch. We need to count only occurrences of '1' in the Hamming mask and examine whether their total number is equal to or less than the user-defined Hamming distance threshold. If so, the mapping is considered to be valid and the read passes the filter. Similarly, if the total number of '1' is greater than the Hamming distance threshold then we cannot be certain whether this is because of the high number of substitutions, or there exist insertions and/or deletions; hence, we need to follow the rest of our algorithm. Our filter can detect not only substitutions but also insertions and deletions in an efficient way, as we discuss next.

#### 2.3.2 Method 2: Insertion and deletion (indel) detection

Our indel detection algorithm is inspired by the original SHD algorithm presented in (Xin *et al.*, 2015). If the substitution detection rejects an alignment, then GateKeeper checks if an insertion or deletion causes the violation (i.e. high number of edits). Figure 1 illustrates the effect of occurrence of edits on the alignment process. If there are one or more base-substitutions or the alignment is exact matching, the matching and mismatching regions can be accurately determined using Hamming distance. As the substitutions have no effect on the alignment of subsequent bases, the number of edits is equivalent to the number of '1's in the resulting Hamming mask. On the other hand, each insertion and deletion can shift multiple trailing bases and create multiple edits in the Hamming mask. Thus, pairwise comparison (bitwise XOR) between the bases of the read and the reference segment is not sufficient. Our indel detection method identifies whether the alignment locations of a read are valid, by shifting individual bases. We need to perform $E$ incremental shifts to the right direction to detect any read that has $E$ deletions, where $E$ is the edit distance threshold. The right shift process guarantees to cancel the effect of deletion. Similarly, we need to perform $E$ incremental shifts to the left direction to detect any read that has $E$ insertions. As we do not have prior knowledge about whether there exist insertions, or deletions, or both, we need to test for every possible case in our algorithm. Thus, GateKeeper generates $2E$ Hamming masks regardless the source of the edit. Each mask is generated after incrementally shifting the candidate read against the reference and performing pairwise comparison (i.e. bitwise XOR operation). A segment of consecutive matches in the one-step right-shifted mask

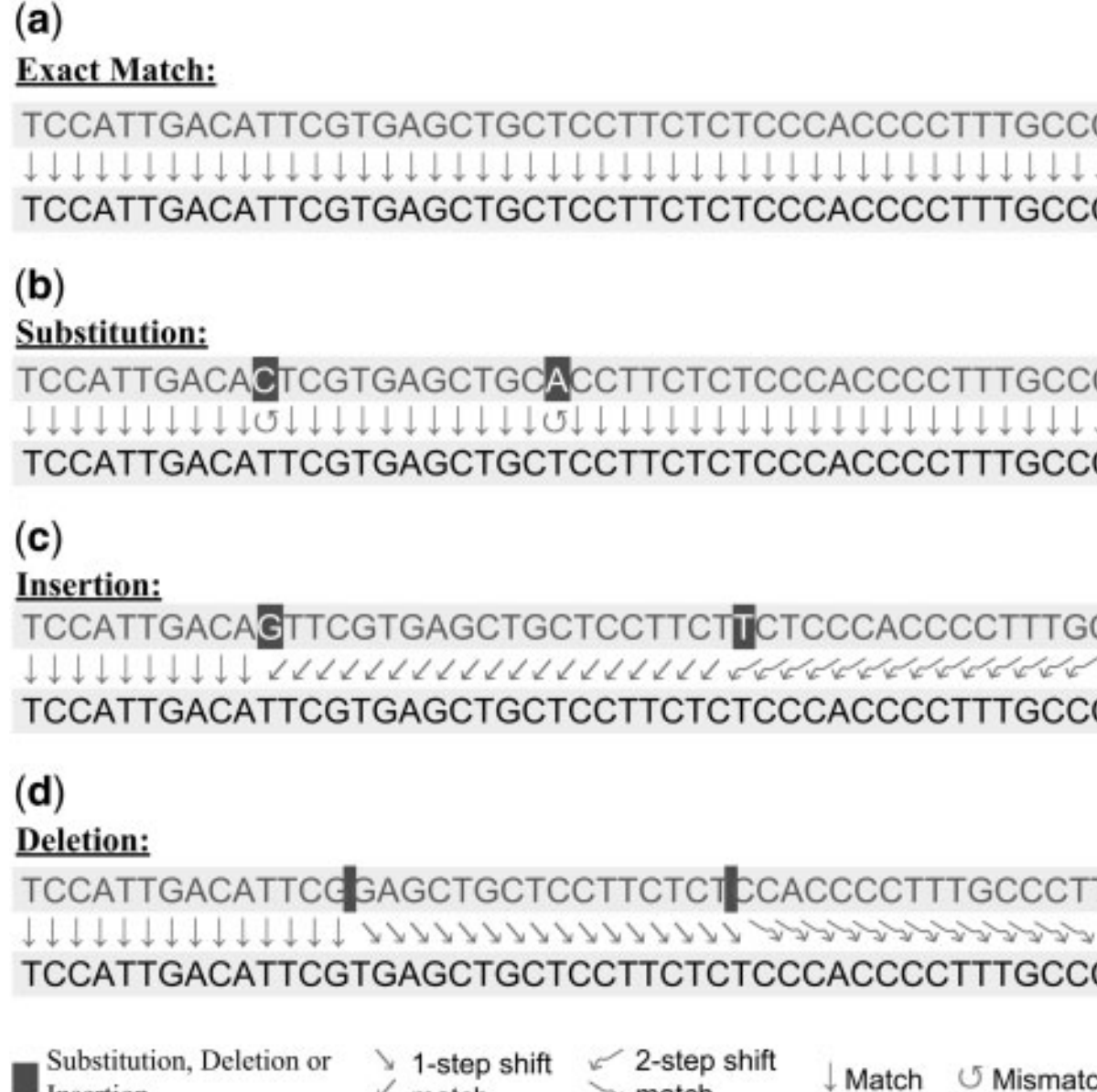

**Fig. 1.** An example showing how various types of edits affect the alignment of two reads. In (**a**) the upper read exactly matches the lower read and thus each base exactly matches the corresponding base in the target read. (**b**) shows base-substitutions that only affect the alignment at their positions. (**c**) and (**d**) demonstrate insertions and deletions, respectively. Each edit has an influence on the alignment of all the subsequent bases

indicates that there is a single deletion that occurred in the read sequence.

Since deletions and insertions affect only the trailing bases, we need to have an additional Hamming mask that is generated with no shifts. This mask helps detect the matches that are located before the first indel. However, this mask is already generated as part of the first method of the algorithm (i.e. Fast Approximate String Matching). The last step is to merge all the $2E+1$ Hamming masks using a bitwise AND operation. This step tells us where the relevant matching and mismatching regions reside in the presence of edits in the read compared to the reference segment. We provide an example of a candidate alignment with all masks that are generated by a single GateKeeper processing core in the Supplementary Material, Figure S9. Identical regions are identified in each shifted Hamming mask as streaks of continuous '0's. As we use a bitwise AND operation, a zero at any position in the $2E+1$ Hamming masks leads to a '0' in the resulting final bit-vector at the same position. Hence, even if some Hamming masks show a mismatch at that position, a zero in some other masks leads to a match ('0') at the same position. This tends to underestimate the actual number of edits and eventually causes some incorrect mappings to pass. To fix this issue, we build a *new hardware-based amending process*. The amending process is first presented in the original SHD filter (Xin *et al.*, 2015) that actually amends (or *flips*) short streaks of '0's (single or double zeros) in each mask into '1's such that they do not mask out '1's in other Hamming masks. Short streaks of '0's do not represent identical sections and thus they are useless. As a result, bit streams such as 101, 1001 are replaced with 111 and 1111, respectively. In SHD, the amending process is accomplished using a 4-bit packed shuffle (SIMD parallel table-lookup instruction), shift and OR operations. The number of computations needed is 4 packed shuffle, $4m$ bitwise

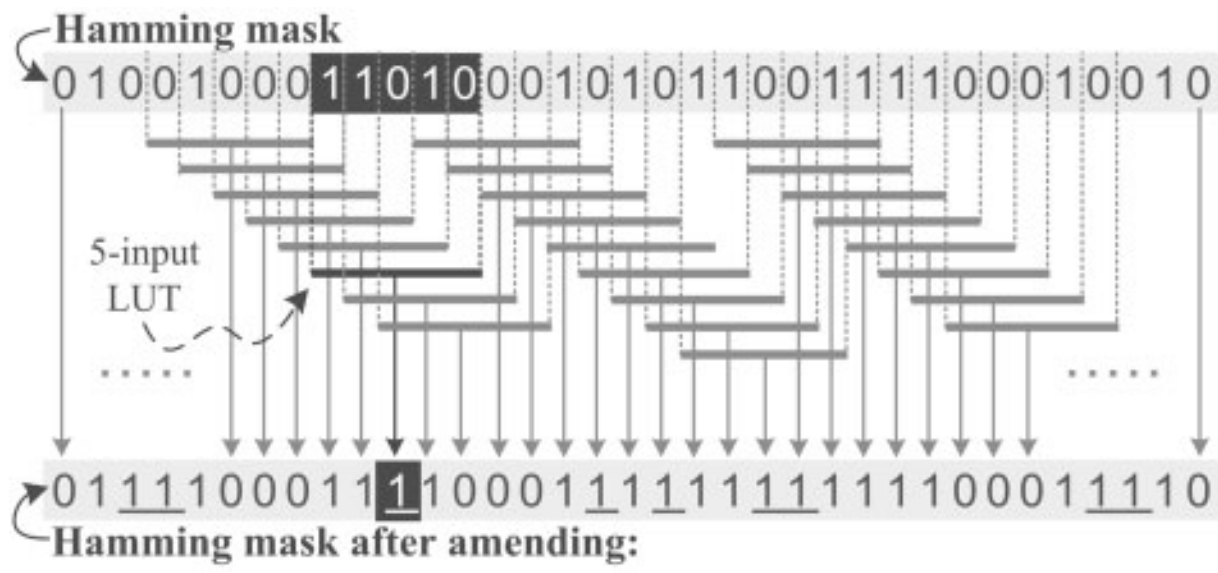

**Fig. 2.** Workflow of the proposed architecture for the parallel amendment operations

OR, and three shift operations for each Hamming mask, which is $(7+4m)(2E+1)$ operations, where $m$ is the read length. We find that this is very inefficient for FPGA implementation. To reduce the number of operations, we propose using dedicated hardware components in FPGA slices. More precisely, rather than shifting the read and then performing packed shuffle to replace patterns of 101 or 1001 to 111 or 1111 respectively, we perform only packed shuffle *independently and concurrently* for each bit of each Hamming mask. As illustrated in Figure 2, the proposed architecture for amendment operations contains one 5-input look-up table (LUT) dedicated for each output bit, except the first and last output bits. We provide full details of our amending architecture in the Supplementary Material (Section 1.3). Using this dedicated architecture, we are able to get rid of the four shifting operations and perform the amending process concurrently for all bits of any Hamming mask. Thus, *the required number of operations is only $(2E+1)$ instead of $(7+4m)(2E+1)$ for a total of $(2E+1)$ Hamming masks*. This saves a considerable amount of the filtering time, reducing it by 407× for a read that is 100 bp long.

### 2.3.3 Method 3: Minimizing hardware resource overheads

The short reads are composed of a string of nucleotides from the DNA alphabet $\sum=\{A, C, G, T\}$. Since the reads are processed in an FPGA platform, the symbols have to be encoded in to a unique binary representation. We need 2 bits ($\log_2|\sum|$ bits) to encode each symbol. Hence encoding a read sequence of length $m$ results in a $2m$-bit word. Encoding the reads into a binary representation introduces overhead to accommodate not only the encoded reads but also the Hamming masks as their lengths also double (i.e. $2m$). The issue introduced by encoding the read can be even worse when we apply certain operations on these Hamming masks. For example, the number of LUTs required for performing the amending process on the Hamming masks will be doubled, mainly due to encoding the read. To reduce the complexity of the subsequent operations on the Hamming masks and save about half of the required amount of FPGA resources, we propose a new solution. We observe that comparing a pair of DNA nucleotides is similar to comparing their binary representations (e.g., comparing A to T is similar to comparing '00' to '11'). Hence, comparing each two bits from the binary representation of the read with their corresponding bits of the reference segment generates a single bit that represents one of two meanings; either match or mismatch between two bases. This is performed by encoding each two bits of the result of the pairwise comparison (i.e. bitwise XOR) into a single bit of '0' or '1' using OR operations in a parallel fashion, as explained in Figure 3. This makes the length of each Hamming mask equivalent to the length of the original read, without affecting the meaning of each bit of the mask. The modified Hamming masks are then merged together in $2E$ bitwise AND operations. Finally, we count the number of ones (i.e. edits) in the final bit-vector mask; if the count is less than the edit distance threshold, the filter accepts the mapping.

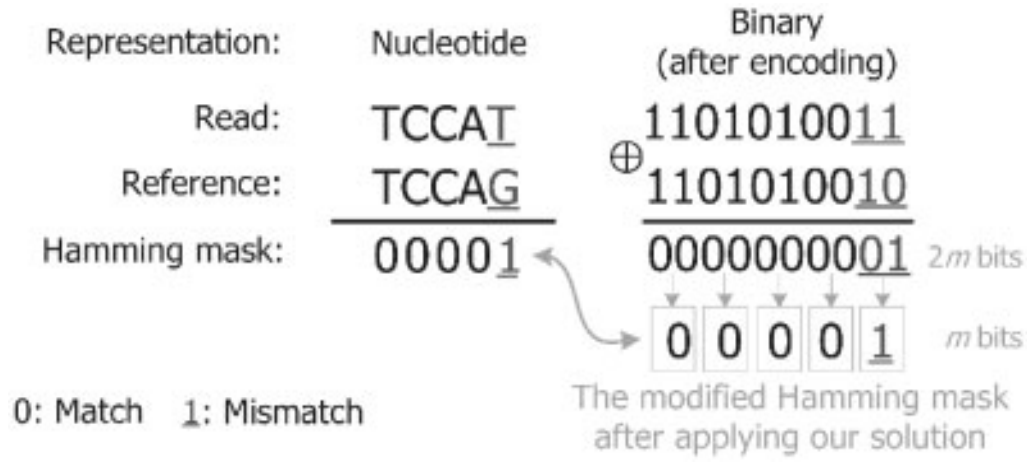

**Fig. 3.** An example of applying our solution for reducing the number of bits of each Hamming mask by half. We use a modified Hamming mask to store the result of applying the bitwise OR operation to each two bits of the Hamming mask. The modified mask maintains the same meaning of the original Hamming mask

**Table 1.** Overall benefits of GateKeeper over SHD in terms of number of operations performed

| # of operations for SHD: | |
|---|---|
| – $m(2E+1)$ bitwise XOR[b]. | – $4m(2E+1)$ bitwise OR.[a] |
| – $2E$ shift. | – $4(2E+1)$ packed shuffle.[a] |
| – $3(2E+1)$ shift.[a] | |
| # of operations for GateKeeper: | |
| For Substitution Detection | For Indel Detection |
| – $2m$ bitwise XOR. | – $2m(2E+1)$ bitwise XOR. |
| | – $2E$ shift. |
| | – $m(2E+1)$ bitwise OR. |
| | – $(2E+1)$ look-up table.[a] |

[a]This operation is required for the amending process.
[b]*E*: edit distance threshold. *m*: read length.

## 2.4 Novelty

GateKeeper is the only read mapping filter that takes advantage of the parallelism offered by FPGA architectures in order to expedite the alignment filtering process. GateKeeper supports both Hamming distance and edit distance in a fast and efficient way. Each GateKeeper processing core performs all operations defined in the GateKeeper algorithm (Supplementary Material, Section 1.3, Algorithm 1). Table 1 summarizes the relative benefits gained by each of the aforementioned optimization methods over the best previous filter, SHD ($E$ is the user-defined edit distance threshold and $m$ is the read length). When a read matches the reference exactly, or with few substitutions, GateKeeper requires *only 2m bitwise XOR operations*, providing substantial speedup compared to SHD, which performs a much greater number of operations. However, this is not the only benefit we gain from our first proposed method (i.e. Fast Approximate String Matching). As this method provides an accurate examination for alignments with only substitutions (i.e. no deletions or insertions), we can directly skip calculating their optimal alignment using the computationally expensive alignment algorithms (i.e. verification step). For more general cases such as deletions and insertions, GateKeeper still requires far fewer operations (as shown in Table 1) than the original SHD filter, due to the optimization methods outlined above. Our improvements over SHD help drastically reduce the execution time of the filtering process. The rejected alignments by our GateKeeper filter are *not* further examined by a verification step. Thus, GateKeeper leads to the acceleration of the entire read mapping process, as our evaluation quantitatively shows (Section 3).

**Table 2.** FPGA resource utilization for a single GateKeeper core

| | Resource utilization % | | | | |
|---|---|---|---|---|---|
| Read length | 100 bp | | 300 bp | | |
| Edit distance | 2 | 5 | 2 | 5 | 15 |
| Slice LUT[a] | 0.39% | 0.71% | 1.27% | 2.2% | 4.82% |
| Slice Register[b] | 0.01% | 0.01% | 0.01% | 0.01% | 0.01% |

[a]LUT: look-up tables.
[b]Flip-flop.

# 3 Evaluation

To implement and evaluate GateKeeper, we use a Xilinx VC709 board (Xilinx, 2014), which features a Virtex-7 XC7VX690T-2FFG1761C FPGA (Xilinx, 2015), and a 3.6 GHz Intel i7-3820 CPU with 8 GB RAM as the host and to run all experiments. We build the FPGA design with Vivado 2014.4 in Verilog. We use RIFFA 2.2 (Jacobsen *et al.*, 2015) to perform the host-FPGA PCIe communication. We configure RIFFA 2.2 as Gen3 4-lane PCIe.

## 3.1 Theoretical speedup

We first examine the maximum speedup theoretically possible with our architecture, assuming the only constraint in the system is the FPGA logic. To this end, we calculate the number of mappings that our accelerator board can potentially examine in parallel using as many GateKeeper processing cores as possible. Table 2 shows the resource utilization of a single processing core for two read lengths of 100 and 300 bp, with different edit distance thresholds. We find that a single processing core for a read length of 300 bp shows 3-fold increase in the number of LUTs compared to its counterpart for a read length of 100 bp, for the same edit distance threshold. This observation is supported by theory: as we show in Table 1, the number of operations of GateKeeper is proportional to both read length and edit distance threshold. Based on the resource report in Table 2, we estimate that we can design GateKeeper, on the VC709 FPGA, to process up to 140 alignments of 100 bp reads and edit distance threshold of up to 5% in parallel in a single clock cycle. The number of alignments drops to 20 for a read length of 300 bp and $E = 15$. The bottleneck in this idealized system is transferring a total of 28 000 (140 alignment × 100 bp × 2 bits for encoding) bits in a single clock cycle into the FPGA, which is not practical for any of the existing PCIe drivers that supply data to the FPGA. For instance, RIFFA (Jacobsen *et al.*, 2015) transmits the mapping pairs into the FPGA in 'packages' of 128 bits per clock cycle at a clock speed of 250 MHz (i.e. 4 nanoseconds). We conclude that the theoretical speedup provided by GateKeeper is extremely large, but practical speedup, which we will examine next, is mainly limited by the data transfer rate into the accelerator.

**Table 3.** Overall system resource utilization under different read lengths and edit distance thresholds

| | Resource utilization % | | | |
|---|---|---|---|---|
| Read length | 100 bp | | 300 bp | |
| Edit distance | 2 | 5 | 2 | 15 |
| Slice LUT | 32% | 45% | 50% | 69% |
| Slice register | 2% | 2% | 17% | 91% |
| Block memory | 2% | 2% | 2% | 2% |

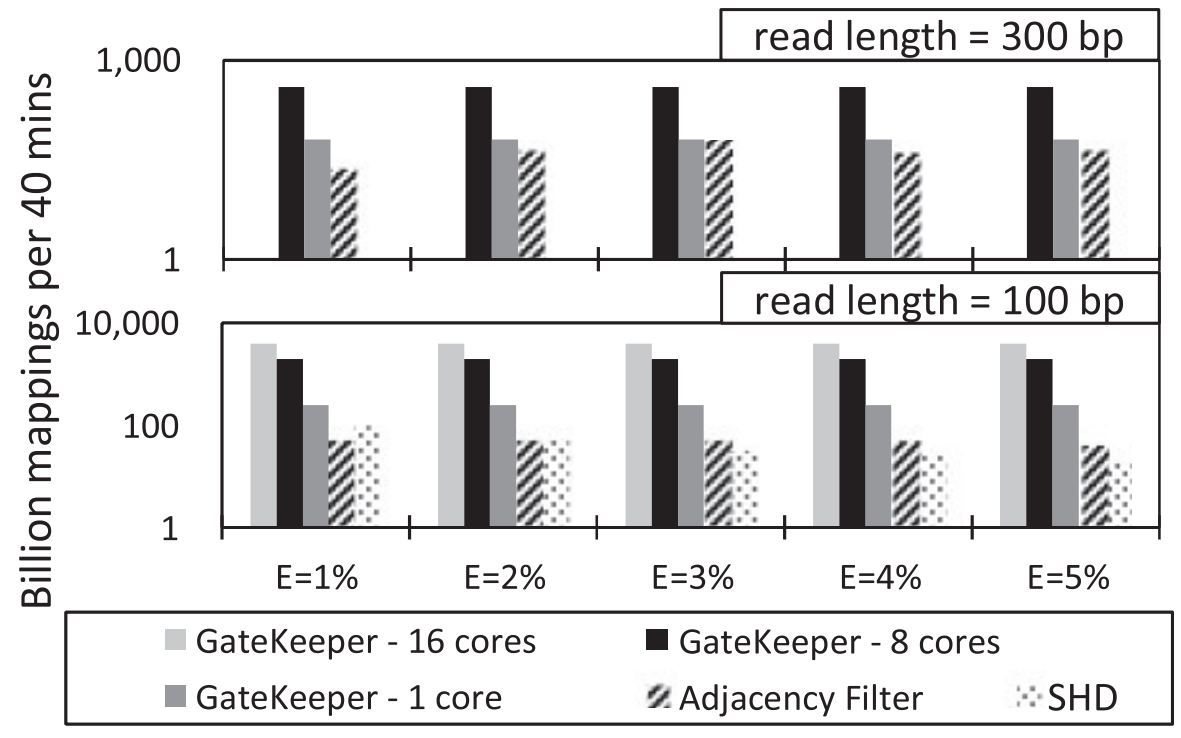

**Fig. 4.** Performance of GateKeeper, SHD, and the Adjacency Filter in terms of the number of examined mappings across different edit distance thresholds and read lengths. The y-axis is on a logarithmic scale. SHD does not support 300 bp long reads

## 3.2 Experimental speedup

***Throughput and resource analysis.*** Filtering speed of GateKeeper is dependent on the total number of concurrent processing cores and the clock frequency. The number of processing cores is determined by the maximum data throughput and the available FPGA resources. The operating frequency of the accelerator is 250 MHz. At this frequency, we observe a data throughput of nearly 3.3 GB/s, which corresponds to ∼13.3 billion bases per second, nearly reaching the maximum throughput of 3.64 GB/s provided by the RIFFA communication channel that feeds data into the FPGA (Jacobsen *et al.*, 2015). Table 3 lists the resource utilization of the entire design including the PCIe communication logic, for various read lengths and edit distance thresholds. For a read length of 100 bp, we find that we can align each read against up to 16 different reference segments in parallel, without violating the timing constraints (e.g. maximum operating frequency). This design occupies about 50% of the available FPGA resources (i.e. slice LUTs). We find that as read length increases, timing constraints of the design can be violated. By pipelining the design (i.e. shortening the critical path delay of each processing core by dividing it into stages or smaller tasks), we can meet the timing constraints and achieve more parallelism. However, pipelining the design comes with the expense of increased register utilization. For a read length of 300 bp, GateKeeper can process up to 8 alignments concurrently and use 91% of the available registers. As our design is FPGA-platform independent, FPGAs with higher logic density (such as Xilinx UltraScale+ FPGAs) can be used, to achieve more parallelism and higher data throughput. Next, we evaluate the effect of varying the number of processing cores on the execution time of GateKeeper.

***Speedup versus existing filters.*** We now evaluate the execution time of GateKeeper compared to the best existing filters. We use mrFAST (Alkan *et al.*, 2009) mapper to retrieve all potential mappings (read-reference pairs) from two datasets. The first set (ERR240727_1) contains about 4 million real reads, each of length 100 bp, from the 1000 Genomes Project Phase I (Consortium, 2012). The second set contains about 100 thousand reads, each of length 300 bp, simulated from the human genome using the *mason* simulator (http://packages.seqan.de/mason/). Figure 4 shows the number of mappings that are processed by GateKeeper (with different numbers of processing cores), SHD, and the Adjacency Filter within 40 minutes. To ensure as fair a comparison as possible, we evaluate Gate Keeper using a single FPGA chip and run both SHD and the Adjacency Filter using a single CPU core. We believe our comparison is fair because we compare GateKeeper running on a part of a single FPGA chip to SHD/Adjacency-Filter running on a part of a single CPU (Section 1.5, Supplementary Material). Both SHD and the Adjacency Filter are software filters (i.e. cannot run on an FPGA) and they do not support multithreading. SHD supports a read length up to only 128 bp (due to SIMD registers size). Under different edit distance thresholds (up to 5% of the read length), GateKeeper provides consistently good performance.

On average, GateKeeper for 100 bp reads is 130x faster than SHD and 90× faster than the Adjacency Filter. For longer reads (i.e. 300 bp), GateKeeper is also, on average, 10× faster than the Adjacency Filter. As edit distance threshold increases, Gatekeeper's speedup over SHD and the Adjacency Filter also increases (e.g. up to 105× and 215× faster than the Adjacency Filter and SHD, respectively, when $E = 5$ edits and read length = 100 bp). This is because our architecture offers the ability to perform all computations in a parallel fashion (as we explained when we described our three new methods in the GateKeeper core). Note that the Adjacency Filter becomes faster than SHD as E increases, but at the expense of accuracy, as we will show soon. We conclude that GateKeeper greatly improves the performance of alignment filtering by at least one order of magnitude. GateKeeper also scales very well over a wide range of both edit distance thresholds and read lengths.

## 3.3 Filtering accuracy

An ideal filter should be both fast and accurate in rejecting the incorrect mappings. We evaluate the accuracy of GateKeeper by computing its true negative, false positive and false negative rates. We use the Needleman-Wunsch algorithm to benchmark the three filters as this algorithm has both zero false positive and zero false negative rates. To evaluate the accuracy of SHD regardless of the limitation of its SIMD implementation (i.e. limited read length), we implement SHD in C and refer to it as SHD-C. We also compare the accuracy of our filter with SHD and the Adjacency Filter using both simulated and real mapping pairs. We simulate reads from the human genome using the *mason* simulator. The configuration and parameters used in our experiment are provided in Supplementary Material (Section 1.4). We generate five sets, each of which contains 400 000 Illumina-like reads. Each set has an equal number of reads of length 64, 100, 150 and 300 bp. While two sets have a low number of different types of edits, the other three sets have a high number of substitutions, insertions and deletions. The purpose of simulating the low-edit reads is that we want most of the reads to have edits less than the allowed threshold. This enables us to quantify the false negatives (i.e. correct mappings that are rejected by the filter) of the three filters with different read lengths. On the other hand, we use the edit-rich reads to evaluate the robustness of the three filters to incorrect mappings. This enables us to quantify both the false positives and true negatives. While the false positive rate is the rate of incorrect mappings that are accepted by the filter, the true negative rate is the rate of incorrect mappings that are rejected by the filter. Figure 5(a) shows the result of this experiment. We also consider a more realistic scenario in which reads can have a combination of substitutions and indels. Instead of simulated reads, we use the first 30 million pairs produced by mrFAST when the dataset ERR240727_1 mapped to the human genome to evaluate both the false positive and true negative rates of the three filters, as shown in Figure 5(b).

Based on these results, we make five main observations. (i) Using the low-edit reads, we observe that the three filters *never filter out correct mappings*; hence, they provide a lossless filtering mechanism with a false negative rate of zero. (ii) We find that GateKeeper is very effective and superior to the Adjacency Filter at both substitution and indel detection. Figure 5(a) shows the average false positive and true negative rates of the three filters, respectively, using the three simulated edit-rich sets. We observe that both GateKeeper and SHD have the same false positive and true negative rates. (iii) On average, GateKeeper produces a false positive rate of 4%, which is much smaller (on average, 0.25×) than that of the Adjacency Filter. (iv) GateKeeper rejects a significant fraction of incorrect mappings (e.g. 84% to 99.9% of the mappings, depending on the edit distance threshold used) and thus avoids expensive verification computations required by alignment algorithms. GateKeeper rejects up to 20% more incorrect mappings than the Adjacency Filter. (v) The Adjacency Filter is more robust in handling indels than in handling substitutions. This is expected as the presence of one or more substitutions in any seed is counted by the Adjacency Filter as a single mismatch. The effectiveness of the Adjacency Filter for substitutions and indels diminishes when $E$ becomes larger than 3%. The detailed results for each of the three edit-rich sets are provided in the Supplementary Material (Section 1.4). We conclude that Gatekeeper's accuracy is as good as that of the best previous filter, SHD, and much better than that of the Adjacency Filter yet GateKeeper is much faster than both SHD and the Adjacency Filter (as we showed earlier). Hence, GateKeeper is extremely fast and accurate.

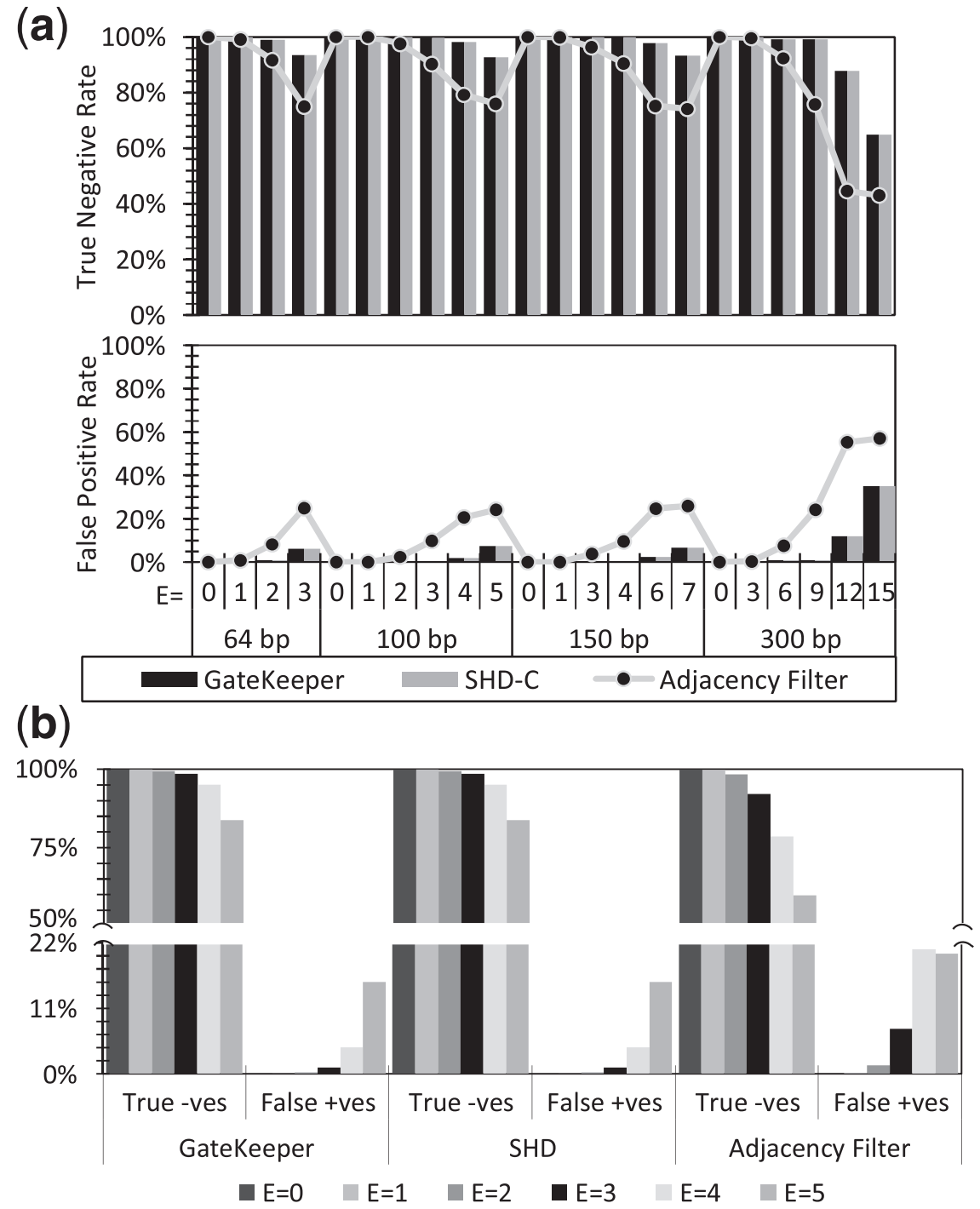

**Fig. 5.** Accuracy of GateKeeper, SHD and the Adjacency Filter across different edit distance thresholds (E) and read lengths. We calculate the false positive [False +ves] and true negative [True -ves] rates using (**a**) simulated and (**b**) real mapping pairs

## 3.4 Verification

GateKeeper is a standalone filter and can be integrated with any existing reference-based mapper. GateKeeper does *not* replace the local/global alignment algorithms (e.g. Smith–Waterman (Smith and Waterman, 1981) and Needleman–Wunsch (Needleman and Wunsch, 1970)). GateKeeper should be followed by an alignment verification step, which precisely verifies the alignments that pass our filter and eliminates the false positives (as provided in the Supplementary Material, Fig. S9). The verification step is accurate and admits zero false positive rate. It also allows specifying a cost to each edit (i.e. a scoring system). Such integration is mapper-specific and will be explored in detail for various mappers in our future research. In this work, we mainly focus on and deeply evaluate the benefits and downsides of our filtering algorithm and architecture independently of any mapper it can be combined with. Nonetheless, we have a preliminary assessment on the overall benefits of integrating GateKeeper with the mrFAST mapper (Alkan *et al.*, 2009). We select mrFAST for two main reasons. (i) It already includes the Adjacency Filter (Xin *et al.*, 2013) as a pre-alignment step, so it constitutes a state-of-the-art baseline. (ii) It utilizes a banded Levenshtein edit distance algorithm (Ukkonen, 1985) that is parallelized using the Intel SSE instructions, and thus it utilizes the capabilities of state-of-the-art hardware. Table 4 summarizes the effect of pre-alignment on the overall mapping time, when all reads from ERR240727_1 (100 bp) and Set_5 (300 bp, mason-simulated deletion-rich reads) are mapped to the human genome with an edit distance threshold of 5%. We make three observations. (i) GateKeeper is at least 41 times faster than the banded dynamic programming alignment algorithm (Ukkonen, 1985). (ii) The verification time drops by a factor of 10 after replacing the Adjacency Filter with GateKeeper as the pre-alignment step. (iii) GateKeeper reduces the overall mapping time of mrFAST (mrFAST-2.6) by a factor of 1.3–3. Details are provided in the Supplementary Material, Section 1.6.

**Table 4.** Overall mrFAST mapping time (in hours) with and without a pre-alignment step, with an edit distance threshold of 5%

| Read length/E | mrFAST version/pre-alignment type | Filtering & verification time (speed-up) | Overall mapping time (speed-up) |
|---|---|---|---|
| 100 bp /5 edits | 2.1/No Pre-alignment | 22.60 h (1×) | 24.27 h (1×) |
| | 2.6/Adjacency Filter | 5.65 h (4×) | 7.31 h (3.3×) |
| | 2.1/GateKeeper | 0.55 h (41×) | 2.50 h (9.7×) |
| 300 bp /15 edits | 2.1/No Pre-alignment | 0.94 h (1×) | 1.02 h (1×) |
| | 2.6/Adjacency Filter | 0.04 h (24×) | 0.12 h (8×) |
| | 2.1/GateKeeper | 0.003 h (279×) | 0.09 h (11×) |

## 4 Future work

GateKeeper shows that there is a great benefit in designing an alignment filtering accelerator to handle the flood of sequenced data. Since a single-core GateKeeper has only a small footprint on the FPGA, we can combine our architecture with any of the FPGA-based accelerators for BWT-FM or hash-based mapping techniques on a single FPGA chip. With such a combination, the end result would be an efficient and fast multi-layer mapping system: alignments that pass GateKeeper can be further verified using a dynamic programing based alignment algorithm *within* the same chip. We leave this combination for future work. Another potential target of our research is to influence the design of more intelligent and attractive sequencing machines by integrating GateKeeper inside them, to perform real-time pre-alignment. This approach has two benefits. First, it can hide the complexity and details of the underlying hardware from users who are not necessarily fluent in FPGAs (e.g. biologists and mathematicians). Second, it allows a significant reduction in total genome analysis time by starting read mapping while still sequencing (Lindner *et al.*, 2016). Our next efforts will also focus on investigating the sources of the false positives and explore the possibility of eliminating them to achieve a dynamic-programming-free alignment approach or a more accurate filter.

## 5 Summary

In this paper, we propose the first hardware accelerator architecture for pre-alignment in genome read mapping. In our experiments, GateKeeper can filter up to ~4 trillion mappings within 40 mins using a single FPGA chip while preserving all correct ones. Comparison against the best two software-based alignment filters reveals the following: (i) Our filter provides, on average, 90-fold and 130-fold speedup compared to the Adjacency Filter and SHD, respectively. (ii) Our filter is as accurate as the SHD and 4 times more accurate than the Adjacency Filter. We conclude that GateKeeper is both a fast and an accurate filter that can improve the performance of existing and future read mappers. Our preliminary results show that the addition of GateKeeper as the pre-alignment step can reduce the filtering and verification time of the mrFAST mapper by a factor of 10.

Our design is open source and freely available online. To our knowledge, GateKeeper is the first open-source FPGA-based alignment filtering accelerator for genome analysis. As such, we hope that it catalyzes the development and adoption of such hardware accelerators in genome sequence analysis, which are becoming increasingly necessary to cope with the processing requirements of greatly increasing amounts of genomic data.

## Funding

This study is supported by NIH Grant (R01 HG006004 to O. Mutlu and C. Alkan) and a Marie Curie Career Integration Grant (PCIG-2011-303772) to C. Alkan under the Seventh Framework Programme. M. Alser also acknowledges support from the Scientific and Technological Research Council of Turkey, under the TUBITAK 2215 program.

*Conflict of Interest*: none declared.

# 1 Supplementary Materials

## 1.1 Read Mappers

Short read mappers typically fall into one of two main categories (Canzar and Salzberg, 2015): (1) Burrows-Wheeler Transformation and Ferragina-Manzini index (BWT-FM)-based methods and (2) Seed-and-extend based methods. Both types have different strengths and weaknesses. The first approach (implemented by BWA (Li and Durbin, 2009), BWT-SW (Li, et al., 2004), Bowtie (Langmead, et al., 2009) and Bowtie2 (Langmead and Salzberg, 2012)) is efficient at finding the *best* mappings of a read (i.e., the mappings with the fewest number of edits), and hence we refer to them as *best-mappers*. Mappers in this category use aggressive algorithms to optimize the candidate location pools to find closest matches, and therefore may not find many potentially-correct mappings (Firtina and Alkan, 2016). Their performance degrades as either the sequencing error rate increases or the genetic differences between the subject and the reference genome are more likely to occur (Li and Durbin, 2009). This is due to the nature of BWT-FM as it entails a *global alignment* (i.e., alignment from the first base to the last one) with respect to the sequenced reads. The second approach, *seed-and-extend mappers* (also referred to as *hash-based* mappers), such as FastHASH (Xin, et al., 2013), mrFAST/mrsFAST (Alkan, et al., 2009; Hach, et al., 2010), SHRiMP2 (David, et al., 2011), and BFAST (Homer, et al., 2009), build very comprehensive but overly large candidate location pools and rely on filters and *local alignment* techniques to remove incorrect mappings from consideration in the verification step. Mappers in this category are able to find *all correct mappings* of a read, but waste computational resources for identifying and rejecting *incorrect* mappings. As a result, they are slower than BWT-FM-based mappers. A recent work (Yorukoglu, et al., 2016) shows that by removing redundancies in the reference genome and also across the reads, seed-and-extend mappers can be faster than BWT-FM-based mappers. A hybrid method that incorporates the advantages of each approach can be also utilized for long read alignment (i.e., up to few million bases), such as BWA-MEM (Li, 2013).

## 1.2 Accelerating Read Mappers

A majority of read mappers are based on machines equipped with general-purpose central processing units (CPUs). While the HTS platforms generate half a trillion bp per day, the state-of-the-art CPU-based read mappers can align only a few billion of them against the human genome (Langmead, et al., 2009; Li, 2013). As long as the gap between the CPU computing speed and the very large amount of sequenced data widens, CPU-based mappers become less favorable due to their limitations in accessing data (Arram, et al., 2013; Houtgast, et al., 2015; Liu, et al., 2012; Luo, et al., 2013; Olson, et al., 2012; Waidyasooriya, et al., 2014). To tackle this challenge, many attempts were made to accelerate the operations of read mapping. Most existing works can be divided into two main approaches: (1) designing hardware accelerators, (2) developing efficient alignment filters.

### 1.2.1 Using Hardware Accelerators

Hardware accelerators for read mapping are becoming increasingly popular as viable solutions for expediting the operations of existing mappers using various new processing platforms, such as GPUs (Benkrid, et al., 2012; Liu, et al., 2012; Luo, et al., 2013) and FPGAs (Benkrid, et al., 2012; Chen, et al., 2013; Houtgast, et al., 2015; Luo, et al., 2013; Olson, et al., 2012; Sogabe and Maruyama, 2013; Sogabe and Maruyama, 2014; Waidyasooriya and Hariyama, 2015; Waidyasooriya, et al., 2014). Fig. 6 illustrates the existing read mappers implemented in various platforms. FPGA acceleration platforms seem to yield the highest performance gain (Aluru and Jammula, 2014; Arram, et al., 2013; Olson, et al., 2012; Waidyasooriya and Hariyama, 2015; Waidyasooriya, et al., 2014), especially for applications with unpredictable and highly irregular memory access patterns such as BWT-based search, which poses difficult challenges for the efficient implementation in CPUs or GPUs (Waidyasooriya and Hariyama, 2015). FPGA-based read mappers often demonstrate one to two orders of magnitude speedups against their GPU-based counterparts (Arram, et al., 2013; Sogabe and Maruyama, 2013; Sogabe and Maruyama, 2014).

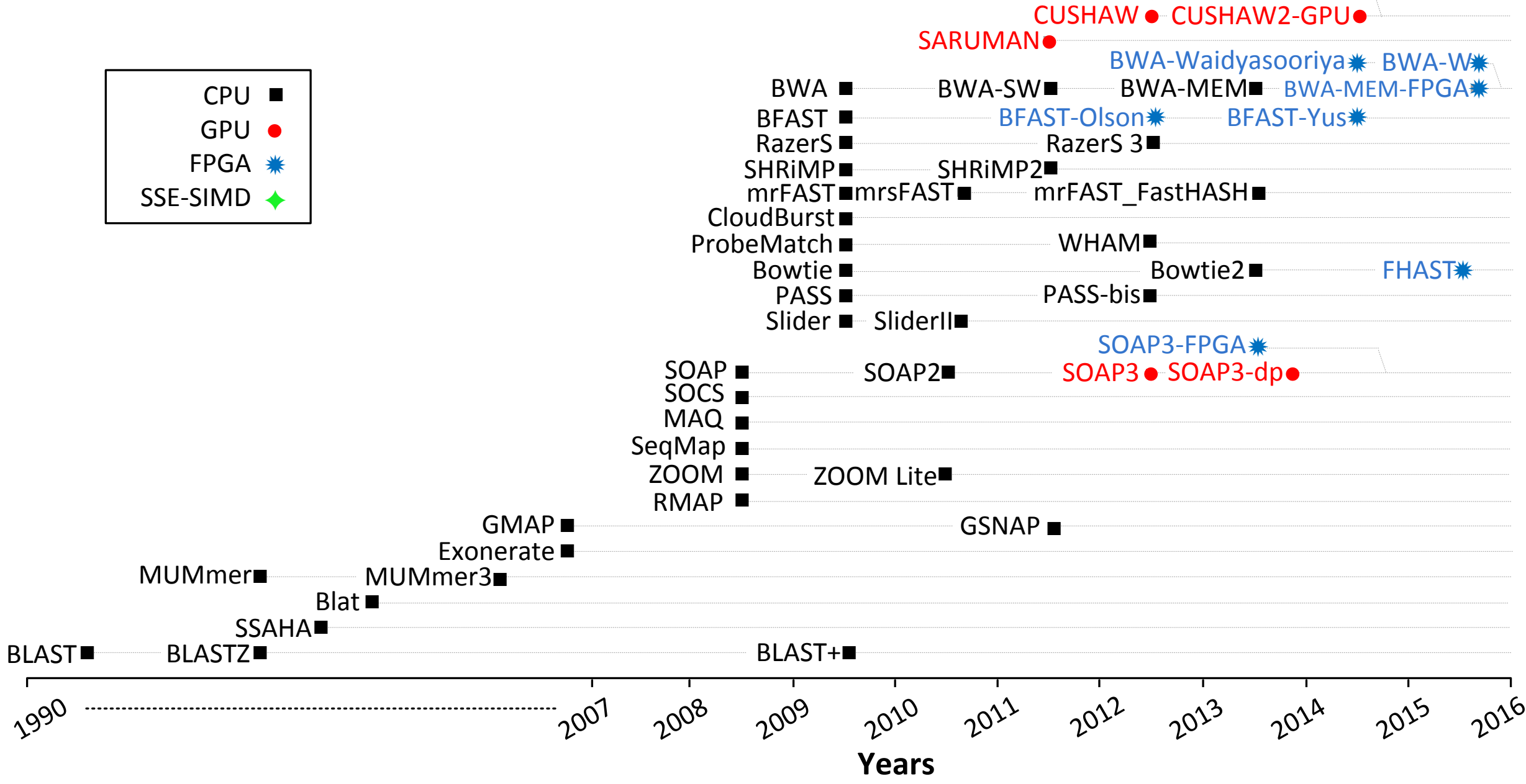

**Fig. 6: Timeline of read mappers. CPU-based mappers are plotted in black, GPU accelerated mappers in red, FPGA accelerated mappers in blue and SSE-based mappers in green. Grey dotted lines connect related mappers (extensions or new versions). The names in the timeline are exactly as they appear in publications, except: SOAP3-FPGA (Arram, et al., 2013), BWA-MEM-FPGA (Houtgast, et al., 2015), BFAST-Olson (Olson, et al., 2012), BFAST-Yus (Sogabe and Maruyama, 2014), BWA-Waidyasooriya (Waidyasooriya, et al., 2014), and BWA-W (Waidyasooriya and Hariyama, 2015).**

Past works used hardware platforms to only accelerate the dynamic programming algorithms (e.g., Smith-Waterman algorithm), as these algorithms contributed significantly to the overall running time of read mappers (Benkrid, et al., 2012; Guerdoux-Jamet and Lavenier, 1995; Szalkowski, et al., 2008). Recently, researchers started paying more attention to integrate the FPGA-accelerated Smith-Waterman algorithm into big-data computing frameworks -such as Apache Spark- for accelerating BWA-MEM (Li, 2013). By this integration, Chen et al. (Chen, et al., 2016) achieve 2.6x speedup over a cloud-based implementation (Chen, et al., 2015) (without FPGA acceleration) of the same mapper. Benkrid et al. (Benkrid, et al., 2012) compare the Smith-Waterman method implemented on the FPGA, GPU, Cell BE, and CPU platforms. The FPGA implementation outperforms all other accelerated implementations, particularly in terms of execution time. FPGAs will likely continue to be the best choice as they enable performing large amounts of computations in a parallel, yet flexible, fashion. Comprehensive surveys on hardware acceleration for computational genomics appeared in (Aluru and Jammula, 2014; Canzar and Salzberg, 2015; Hatem, et al., 2013). Note that there is no work on hardware acceleration of alignment filtering techniques, which we discuss next.

### 1.2.2 Using Alignment Filtering Techniques

The second approach to accelerate read mapping is to incorporate a filtering technique within the read mapper, before the verification step. This filter is responsible for quickly excluding incorrect mappings in an early stage (i.e., as a pre-alignment step) to reduce the number of locations that must be verified via dynamic programming. Existing filtering techniques include three major types of filters.

***Hamming Distance.*** Hamming distance (Hamming, 1950; Ukkonen, 1992) between a pair of sequences is defined as the number of positions at which the corresponding symbols are different. As such, Hamming distance measures the pairwise differences between sequences of equal length. Hence, it can only find substitutions. Calculation of Hamming distance is relatively easy and supported by various hardware platforms. Bitwise operations (mainly XORs) between an appropriate binary representation of a read and a genomic-region can be employed to obtain a bit vector that indicates edits. The Hamming distance of two strings can be calculated simply through a linear scan counting the number of pairwise differences. The key drawback of this filtering approach is that Hamming distance *cannot* correctly find insertions and deletions (indels). Allowing indels is important because both sequencing errors and genetic variations can result in the deletion/insertion of bases and the chance of this happening increases as reads become longer. However, finding indels is computationally more challenging. mrsFAST (Hach, et al., 2010) and mrsFAST-Ultra (Hach, et al., 2014) are examples of read mappers that use Hamming distance as a filtering strategy.

***Seed Filtering.*** Filtering the incorrect mappings using seeds is the basic principle of nearly all seed-and-extend mappers (such as, GEM (Marco-Sola, et al., 2012), RaserS (Weese, et al., 2009), GASSST (Rizk and Lavenier, 2010), RazerS 3 (Weese, et al., 2012), Hobbes (Ahmadi, et al., 2012), FastHASH (Xin, et al., 2013), BitMapper (Cheng, et al., 2015)). Instead of considering one long string as a whole, a seed filter examines all of the string's possible substrings of length q (which are called seeds). Seed filtering is based on the observation that if two sequences are potentially similar, then they share a certain number of seeds. The seed is sometimes called a q-gram or a k-mer. Seeds are used as indices into the reference genome to reduce the search space and speed up the mapping process. The performance and accuracy of seed-and-extend mappers depend on how the seeds are selected in the first stage. Mappers should select a large number of non-overlapping seeds while keeping each seed as infrequent as possible (Kiełbasa, et al., 2011; Xin, et al., 2013; Xin, et al., 2015). There is also a significant advantage to selecting seeds with unequal lengths, as possible seeds of equal lengths can have drastically different levels of frequencies. Finding the optimal set of seeds from read sequences is challenging and complex, primarily because the associated search space is large and it grows exponentially as the number of seeds increases. There are other variants of seed filtering based on the pigeonhole principle (Cheng, et al., 2015; Weese, et al., 2012), non-overlapping seeds (Xin, et al., 2013), gapped seeds (Egidi and Manzini, 2013; Rizk and Lavenier, 2010), variable-length seeds (Xin, et al., 2015), or random permutation of subsequences (Lederman, 2013). We select the Adjacency Filter (AF) from FastHASH (Xin, et al., 2013) as a representative example.

***Shifted Hamming Distance (SHD).*** Another recent filtering technique is called *Shifted Hamming Distance* (SHD) (Xin, et al., 2015). It is based on the pigeonhole principle. That is, if $E$ items are put into $E+1$ boxes, then one or more boxes would be empty. It can be applied in context of sequence alignment as follows: if two reads differ by $E$ edits, then they should share at least a single identical section (free of edits) among $E+1$ non-overlapping sections, where $E$ is the threshold of edit distance. This is due to the fact that the $E$ mismatches would result in dividing the read into $E+1$ identical sections in accordance with their correspondences in the reference. The more edits are involved between a read and the reference, the less contiguous stretches of exact matches they share. SHD relies on identifying these $E+1$ identical sections, between the read and the reference, as a proxy for the edit distance calculations. If there are no more than $E$ edits between the read and the reference, then each non-erroneous segment in the read can be matched to its corresponding region in the reference within $E$ shifts from its position to the right or left direction. The shifting process is inevitable in order to skip the erroneous bases (especially in case of insertions and deletions, as explained in Fig. 1 in the main manuscript). Our crucial observation is that SHD examines each mapping, throughout the filtering process, by performing expensive computations unnecessarily; as SHD uses the same amount of computation regardless the type of edit. In particular, substitutions can be directly measured by the Hamming distance and does not require the shifting process. Thus, SHD is *not* suitable for applications that aim at finding the optimal alignment of reads against a reference where indels are *not* allowed, for example, identifying single nucleotide polymorphisms (SNPs) that are associated with diseases (Hach, et al., 2014); aiming at discovering and developing efficient drugs.

***Conclusions.*** While Hamming distance simply counts the number of substitutions implied by mismatched symbols of two strings at the same position, edit distance additionally accounts for inserted or deleted symbols in one string with respect to the other. Calculating Hamming distance is relatively easy and has been well solved. On the other hand, calculating edit distance efficiently is still the focus of ongoing research. FastHASH has high false positive rates for edit distance thresholds higher than 3 edits (Xin, et al., 2015). On average, as shown experimentally in (Xin, et al., 2015), SHD requires the same execution time as the Adjacency Filter of FastHASH (seed filtering approach). However, SHD produces far fewer (4X fewer) false positives compared to the Adjacency Filter. Therefore, it is crucial to develop a fast and effective filter that can detect incorrect candidate locations and eliminate them before using computationally costly alignment operations.

### 1.2.3 Comparison between Filtering Techniques and Hardware Accelerators

We provide a comparison of existing filters with other existing hardware accelerated read mappers using various platforms. Table 5 summarizes the results. We report the running time of different tools using 100 bp reads with at most 2 edits (unless otherwise mentioned). To provide a fair comparison across FPGA-based architectures, we report the run time of using only a single FPGA chip. In all these studies we surveyed, FPGAs outperform all other accelerator platforms in terms of run time. The best-performing alignment filter (i.e., the SHD filter) is only 3x faster than the fastest FPGA-based mapper (i.e., the work presented in (Arram, et al., 2013)). An ideal filter should have both high accuracy and low running time to compensate the computation overhead introduced by its filtering technique. **Our goal** in this paper is to design a new fast filtering algorithm (building upon SHD) and a new hardware architecture that accelerates it by taking advantage of the computational capabilities of FPGAs. To our knowledge, this is the first work that takes advantage of novel hardware architectures to accelerate *alignment filtering* techniques.

**Table 5: Alignment performance for various state-of-the-art mappers and filters for 100 bp reads with an edit distance threshold of 2.**

| Year | Ref. | Purpose* | Architecture | Platform | # of alignments in 1 sec |
|---|---|---|---|---|---|
| 2015 | (Xin, et al., 2015) | Pre-alignment | Shifted Hamming Distance | Intel SSE | 18,820,572 |
| 2015 | (Xin, et al., 2015) | Verification | Myers's bit-vector (Döring, et al., 2008) | Intel SSE | 2,146,266 |
| 2015 | (Xin, et al., 2015) | Verification | Smith-Waterman (Szalkowski, et al., 2008) | Intel SSE | 201,783 |
| 2015 | (Waidyasooriya and Hariyama, 2015) | Read Mapper | BWT-FM | FPGA (Stratix5) | *(101 bp)* 86,633 |
| 2014 | (Waidyasooriya, et al., 2014) | Read Mapper | BWT-FM | FPGA (Stratix5) | *(90 bp)* 68,259 |
| 2014 | (Liu and Schmidt, 2014) | Verification | Smith-Waterman | GPU | 4,000 |
| 2014 | (Sogabe and Maruyama, 2014) | Read Mapper | Hash-Based (BFAST) | FPGA (Virtex7) | 316,455 |
| 2013 | (Luo, et al., 2013) | Read Mapper | BWT-FM | GPU | 90,907 |
| 2013 | (Chen, et al., 2013) | Read Mapper | Hash-Based | FPGA (Virtex5) | 22,658 |
| 2013 | (Sogabe and Maruyama, 2013) | Read Mapper | Hash-Based (BFAST) | FPGA (Virtex7) | 80,775 |
| 2013 | (Arram, et al., 2013) | Read Mapper | BWT-FM | FPGA (Virtex6) | *(90 bp)* 5,734,265 |
| 2012 | (Liu, et al., 2012) | Read Mapper | BWT-FM | GPU | 90,900 |
| 2012 | (Olson, et al., 2012) | Read Mapper | Hash-Based (BFAST) | FPGA (Virtex6) | *(76 bp)* 183,823 |
| 2012 | (Benkrid, et al., 2012) | Verification | Smith-Waterman | FPGA (Virtex4) | *(128 bp)* 689,543 |
| | | | | GPU | *(128 bp)* 86,192 |
| | | | | Cell BE | *(128 bp)* 216,327 |
| | | | | CPU | *(128 bp)* 5,253 |

* Although the execution times of filters and mappers are not directly comparable, the comparison highlights the necessity of developing an efficient and fast filter.

## 1.3 GateKeeper FPGA Implementation

In this section, we discuss the algorithmic and implementation details of GateKeeper.

### 1.3.1 Overview

Fig. 7 shows the overall architecture of our FPGA-based accelerator, GateKeeper, which consists of an FPGA engine as an essential component and a CPU. The latter is responsible for acquiring and encoding the short reads and transferring the data to and from the FPGA. The FPGA engine is equipped with PCIe transceivers, Read Controller, Mapping Controller, and a set of processing cores that are responsible for examining the read alignment. The workflow of the accelerator starts with reading a repository of short reads and seed locations. All reads are then converted into their binary representation that can be understood by the FPGA engine. Encoding the reads is a preprocessing step and accomplished through a Read Encoder at the host before transmitting the reads to the FPGA chip. Next, the encoded reads are transmitted and processed in a streaming fashion through the fastest communication medium available on the FPGA board (i.e., PCIe). We set the edit distance threshold to 5% of the read length or less as this threshold is widely used for mapping reads against a reference genome (Ahmadi, et al., 2012; Cheng, et al., 2015; Hatem, et al., 2013; Xin, et al., 2015). The output results are transferred back to the CPU in the same order as the input stream in a streaming fashion and then saved in the repository. We design our system to perform alignment filtering in a *streaming* fashion: the accelerator receives a continual stream of short reads, examines each alignment in parallel with others and returns the decision (i.e., whether the alignment is accepted or rejected) instantaneously upon processing. The pseudocode of our new FPGA-friendly filtering algorithm is shown in Algorithm 1. Fig. 8 presents a flowchart representation of all steps involved in the algorithm. Fig. 9 provides an example of all masks that are generated by a single GateKeeper processing core. Although GateKeeper shows a low number of false positives, we still need the dynamic programming algorithm to perform an accurate verification for any mapping that passes our filter.

### 1.3.2 Read Controller

The Read Controller on the FPGA side is responsible for two main tasks. First, it permanently assigns the first data chunk as a reference sequence for all processing cores. Second, it manages the subsequent data chunks and distributes them to the processing cores. The first processing core receives the first read sequence and the second core receives the second sequence and so on, up to the last core. It iterates the data chunk management task until no more reads are left in the repository.

### 1.3.3 Mapping Controller

Following similar principles as the Read Controller, the Mapping Controller gathers the output results of the processing cores. Both the Read Controller and the Mapping Controller preserve the original order of reads as in the repository (i.e., at the host). This is critical to ensure that each read will receive its own alignment filtering result. The results are transmitted back to the CPU side in a streaming fashion and then saved in the repository.

**Algorithm 1:** GateKeeper filtering algorithm

**Input:** Candidate read bit-vector $r = \{ r_1, r_2 ... r_m \}$, Reference bit-vector $f = \{ f_1, f_2 ... f_m \}$, edit distance threshold $E$
**Output:** Pass (return True if the read passes the GateKeeper filter).
**Functions:** *Amend*: Encodes/amends the masks.
**Pseudocode:**

```
//Calculate Hamming distance first.
HammingMask[2E+1] = r ⊕ f;
AmendedMask[2E+1] = Amend (HammingMask[2E+1]);
e = # of '1's in HammingMask[2E+1] after encoding;
if e ≤ E
    |Pass = True;
else //Generate 2E masks with incremental shift.
    for i = 1 to E do
        HammingMask[i] = (r>>i) ⊕ f;
        AmendedMask[i]= Amend (HammingMask[i]);
        HammingMask[i+E] = (r<<i) ⊕ f;
        AmendedMask[i+E]=
                    Amend(HammingMask[i+E]);
    FinalMask = AND(AmendedMask[1 .... 2E+1];
    i=1; e=0;
    while i < m do //Count the differences.
        case (FinalMask[i, i+1, i+2, i+3]):
        [0101],[0110],[1001],[1010],[1011],[1101]:
        e= e +2;
        [0000]: e = e;     default: e = e+1;    i = i+4;
    if e ≤ E
        |Pass = True;
    else
        |Pass = False;
return Pass;
```

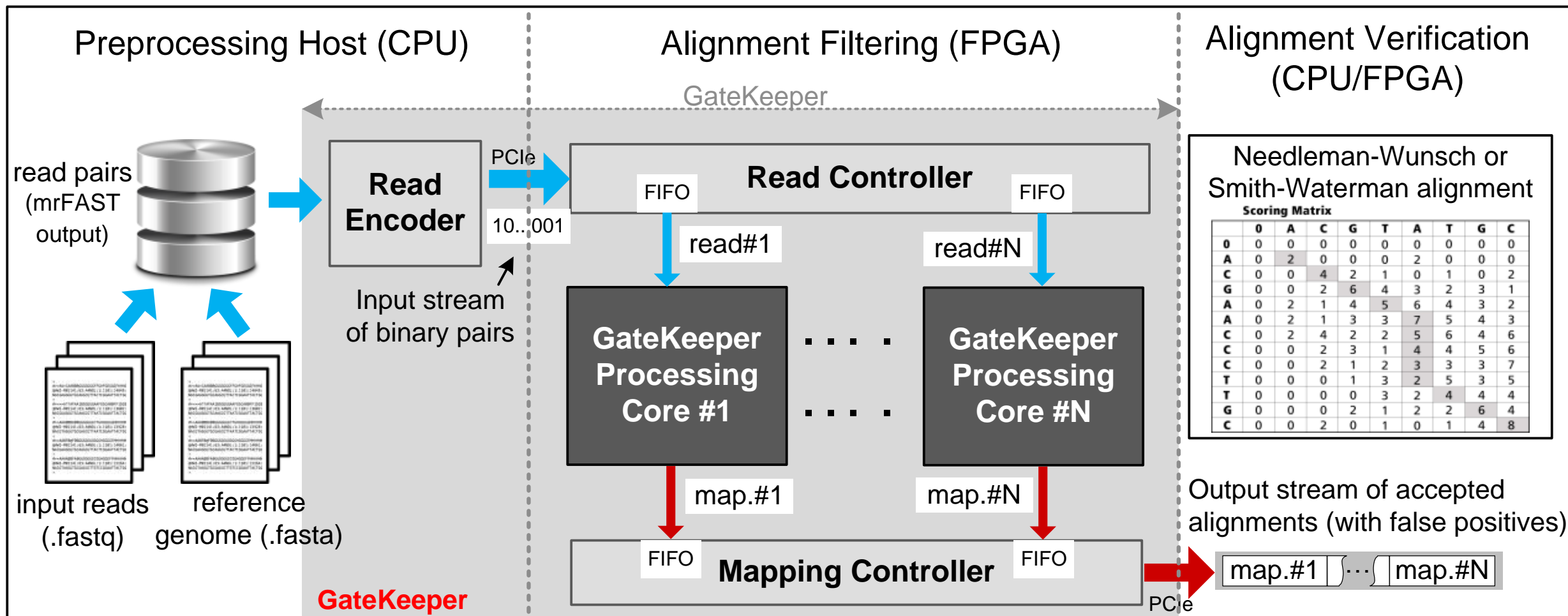

**Fig. 7: Overview of the GateKeeper accelerator architecture.**

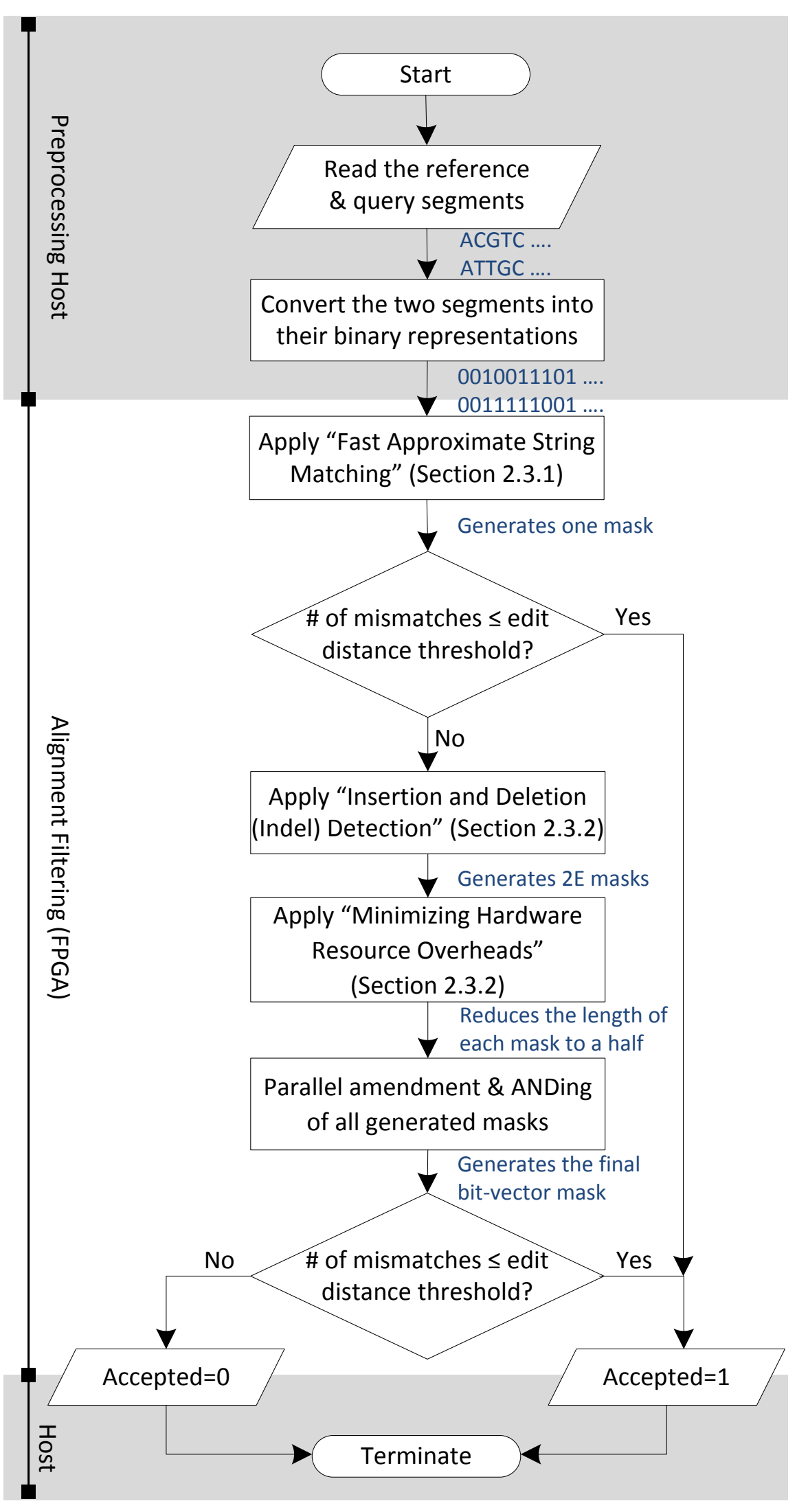

**Fig. 8: A flowchart representation of the GateKeeper algorithm.**

### 1.3.4 New Hardware-Based Amending Process for Supporting Insertion and Deletion Detection

In order to replace all patterns of 101 or 1001 to 111 or 1111 respectively, we use a single 5-input look-up table for each bit of the Hamming mask. The first look-up table copies the bit value of the first input regardless of its value; even if it is zero, it will not be amended as it is not contributing to the 101 or 1001 pattern. Likewise for the last look-up table. Thus, the total number of look-up tables needed is equal to the length of the short read in bases minus 2 for the first and last bases. In each look-up table, we consider a single bit of the Hamming mask and two of its right neighboring bits and two of its left neighboring bits. If the input that corresponds to the output has a bit value of one, then the output copies the value of that input bit (as we only amend zeros). Otherwise, using the previous two bits and the following two bits with respect to the input bit, we can replace any zero of the "101" or "1001" patterns independently from other output bits (details are given in Algorithm 2). All bits of the amended masks are generated at the same time, as the propagation delay through an FPGA look-up table is independent of the implemented function (Xilinx, November 17, 2014). Thus we can process all masks in a parallel fashion without affecting the correctness of the filtering decision.

---

**Algorithm 2:** Amend

---

**Input:** Hamming mask bit-vector, $H = \{ H_1, H_2 ... H_{2m} \}$
**Output:** modified Hamming mask, $A = \{ A_1, A_2 ... A_m \}$
**Pseudocode:**
$i = 0$;
**while** $i < m$ **do** `//Encode Hamming masks`
  $E_i = H_i \mid H_{i+1}$;
  $i = i+2$;
`//Amend 101 and 1001 patterns`
$A_1 = E_1$; `//Initialization`
$A_m = E_m$;
$A_2 = (E_1 \overline{E_2} E_3) \mid (E_1 \overline{E_2}\, \overline{E_3} E_4)$;
$A_{m-1} = (E_{m-2} \overline{E_{m-1}} E_m) \mid (E_{m-3} \overline{E_{m-2}}\, \overline{E_{m-1}} E_m)$;
**for** $i = 3$ **to** $m-2$ **do**
  **if** $E_i == 1$
    $A_i = E_i$;
  **else**
    $A_i = (E_{i-1} \overline{E_i} E_{i+1}) \mid (E_{i-2} \overline{E_{i-1}}\, \overline{E_i} E_{i+1}) \mid (E_{i-1} \overline{E_i}\, \overline{E_{i+1}} E_{i+2})$;
**return** $A$;

---

```
          Query : AAAAAAAAAAAAAAGAGAGAGAGATATTTAGTGTTGCAGCACTACAACACAAAAGAGGACCAACTTACGTGTCTAAAAGGGGGAACATTGTTGGGCCGGA
      Reference : AAAAAAAAAAAAAAGAGAGAGAGATAGTTAGTGTTGCAGCCACTACAACACAAAAGAGGACCAACTTACGTGTCTAAAAGGGGAGACATTGTTGGGCCGG

   Hamming Mask : 0000000000000000000000000010000000000000111111011110001110110101101111111110001000001111011010010101
 1-Deletion Mask : 0000000000000011111111111110011111011111000000000000000000000000000000000000000000011000000000000000
 2-Deletion Mask : 0000000000000010000000001011011100111111111111101111000111011010110111111111000100010011101101001010
 3-Deletion Mask : 0000000000000010111111111110111011001101110111011000100100111111111111100101100110010110111011101111
1-Insertion Mask : 0000000000000111111111111110111110111111011101100010010011111111111110010110011000101011101110111110
2-Insertion Mask : 0000000000001000000000100111110011111111100100011010101001101011111111111110111001111111000111101100
3-Insertion Mask : 0000000000010111111111110111011001100011111111101011011111100110010111011111111011101111010111001000

                          --- Masks after amendment ---

   Hamming Mask : 0000000000000000000000000010000000000000111111111110001111111101111111111110001000001111111111111111
 1-Deletion Mask : 0000000000000011111111111111111111111111000000000000000000000000000000000000000000011000000000000000
 2-Deletion Mask : 0000000000000010000000001111111111111111111111111111000111111111111111111111000100011111111111111110
 3-Deletion Mask : 0000000000000011111111111111111111111111111111111000111111111111111111111111111111111111111111111111
1-Insertion Mask : 0000000000000111111111111111111111111111111111100011111111111111111111111111100011111111111111111110
2-Insertion Mask : 0000000000001000000000111111111111111111111100011111111111111111111111111111111111111100011111111100
3-Insertion Mask : 0000000000011111111111111111111111100011111111111111111111111111111111111111111111111111111111111000

       AND Mask : 0000000000000000000000000010000000000000100000000000000000000000000000000000000000000000100000000000000

Needleman-Wunsch  AAAAAAAAAAAAAAGAGAGAGAGATATTTAGTGTTGCAG-CACTACAACACAAAAGAGGACCAACTTACGTGTCTAAAAGGGGGAACATTGTTGGGCCGG
      Alignment:  |||||||||||||||||||||||||| |||||||||||| |||||||||||||||||||||||||||||||||||||||||||::|||||||||||||||
                  AAAAAAAAAAAAAAGAGAGAGAGATAGTTAGTGTTGCAGCCACTACAACACAAAAGAGGACCAACTTACGTGTCTAAAAGGGGAGACATTGTTGGGCCGG
```

**Fig. 9: An example of an alignment with all its generated masks, where the edit distance threshold is set to 3. The alignment is also examined and compared with the Needleman-Wunsch algorithm. The alignment passes the GateKeeper filter as a false positive, but later gets rejected by the Needleman-Wunsch algorithm.**

### 1.4 mrFAST, SHD, Adjacency Filter, and mason Configurations

In our experiments, we generate the real read set and five simulated sets of short and long Illumina-like reads using the command lines presented in Table 6. The first and third sets have a number of base-substitutions that are drawn from a Gaussian distribution with mean of 3% (of read length) and 16%, respectively. The second, fourth, and fifth sets have a number of indels that are drawn from a Gaussian distribution with a mean of 3% , 16%, and 16%, respectively. Table 6 also provides the command lines used to run both the SHD and the Adjacency Filter. Fig. 10 provides the number of false positives and true negatives of GateKeeper, SHD, and the Adjacency Filter, using simulated reads from three different data sets, namely, Set3, Set4, and Set5.

### 1.5 Ensuring a fair performance comparison

In this section we present our efforts toward ensuring a fair comparison between the hardware and software filters. SHD is a software-based filter that is implemented in C with Intel SSE instructions. Hence, SHD (and similarly the Adjacency Filter) cannot be directly implemented on an FPGA because the architecture of an FPGA is fundamentally different from the architecture of a general-purpose CPU for which SHD is designed. In contrast, GateKeeper is a new filtering mechanism that is designed specifically for an FPGA, and hence it is not a simple alternative implementation of SHD. GateKeeper is a complete system designed for an FPGA, which includes many processing cores. Each processing core is an individual alignment filter. All processing cores work in parallel to achieve an extremely fast filtering speed. Though SHD and the Adjacency Filter do not support multithreading, GateKeeper is designed to utilize the large amounts of parallelism offered by FPGA architectures by integrating as many processing cores as possible in the FPGA chip. These cores are fundamentally different from the CPU cores or threads, as they are extremely small compared to a full CPU core that is used by SHD or the Adjacency Filter. To ensure a fair performance comparison, we use a single FPGA chip to design GateKeeper. In fact, our filter uses part of a single FPGA chip. Likewise SHD and the Adjacency Filter use part of a single CPU chip. Fig. 11 shows the chip layout for both VC709 FPGA and Intel i7 CPU. It shows that GateKeeper (including routing resources, RIFFA logic, and PCIe controller) occupies about 22.5% (406 mm$^2$) of the total chip area. Software that runs on a single CPU core can occupies from 16% (chip area / 6 cores = 378 mm$^2$) up to 77% (1 CPU core + cache + memory controller + IO = 1819 mm$^2$) of the processor chip area. We conclude that GateKeeper occupies a similar amount of chip area or less as SHD and the Adjacency Filter. Note that a "perfectly" fair comparison is extremely difficult because the FPGA architecture is fundamentally different from the CPU architecture. For example, if we hypothetically could increase the number of cores used for SHD/Adjacency-Filter, for a fair comparison, we might want to increase the number of FPGAs used for GateKeeper.

### 1.6 GateKeeper with mrFAST

In this section, we present the overall speedup when taking into account *all mapping steps,* such as generating candidate alignment locations, pre-alignment, verification, and generating the SAM file (i.e., an output file that contains the alignment details, such as the number of edits needed to make the read sequence exactly match the reference segment). We select mrFAST (Alkan, et al., 2009) as an example. However, GateKeeper can be integrated with any mapper that performs sequence alignment for verification. We select mrFAST due to two reasons: (1) It already includes the Adjacency Filter (Xin, et al., 2013) as a pre-alignment step, so it constitutes a state-of-the-art high-performance baseline. (2) It utilizes a banded Levenshtein edit distance algorithm (Ukkonen, 1985) that is parallelized using the Intel SSE instructions, and thus it utilizes the capabilities of state-of-the-art hardware. Table 7 summarizes the total runtime breakdown for mrFAST with and without GateKeeper for short and long (100 and 300 bp) reads.

**Table 6: Read simulator and mapper versions and command lines used in the evaluations.**

| |
|---|
| **Mason: version 0.1.2** |
| **Set 1: (400,000 low-substitution reads)** |
| *mason illumina -N 100000 -i -o Set1_64.fasta -n 64 -f -snN -nN -pmm 0.03 -pi 0 -pd 0 human_g1k_v37.fasta* |
| *mason illumina -N 100000 -i -o Set1_100.fasta -n 100 -f -snN -nN -pmm 0.03 -pi 0 -pd 0 human_g1k_v37.fasta* |
| *mason illumina -N 100000 -i -o Set1_150.fasta -n 150 -f -snN -nN -pmm 0.03 -pi 0 -pd 0 human_g1k_v37.fasta* |
| *mason illumina -N 100000 -i -o Set1_300.fasta -n 300 -f -snN -nN -pmm 0.03 -pi 0 -pd 0 human_g1k_v37.fasta* |
| **Set 2: (400,000 low-indel reads)** |
| *mason illumina -N 100000 -i -o Set2_64.fasta -n 64 -f -snN -nN -pmm 0.01 -pi 0.01 -pd 0.01 human_g1k_v37.fasta* |
| *mason illumina -N 100000 -i -o Set2_100.fasta -n 100 -f -snN -nN -pmm 0.01 -pi 0.01 -pd 0.01 human_g1k_v37.fasta* |
| *mason illumina -N 100000 -i -o Set2_150.fasta -n 150 -f -snN -nN -pmm 0.01 -pi 0.01 -pd 0.01 human_g1k_v37.fasta* |
| *mason illumina -N 100000 -i -o Set2_300.fasta -n 300 -f -snN -nN -pmm 0.01 -pi 0.01 -pd 0.01 human_g1k_v37.fasta* |
| **Set 3: (400,000 substitution-rich reads)** |
| *mason illumina -N 100000 -i -o Set3_64.fasta -n 64 -f -snN -nN -pmm 0.16 -pi 0 -pd 0 human_g1k_v37.fasta* |
| *mason illumina -N 100000 -i -o Set3_100.fasta -n 100 -f -snN -nN -pmm 0.16 -pi 0 -pd 0 human_g1k_v37.fasta* |
| *mason illumina -N 100000 -i -o Set3_150.fasta -n 150 -f -snN -nN -pmm 0.16 -pi 0 -pd 0 human_g1k_v37.fasta* |
| *mason illumina -N 100000 -i -o Set3_300.fasta -n 300 -f -snN -nN -pmm 0.16 -pi 0 -pd 0 human_g1k_v37.fasta* |
| **Set 4: (400,000 insertion-rich reads)** |
| *mason illumina -N 100000 -i -o Set4_64.fasta -n 64 -f -snN -nN -pmm 0 -pi 0.16 -pd 0 human_g1k_v37.fasta* |
| *mason illumina -N 100000 -i -o Set4_100.fasta -n 100 -f -snN -nN -pmm 0 -pi 0.16 -pd 0 human_g1k_v37.fasta* |
| *mason illumina -N 100000 -i -o Set4_150.fasta -n 150 -f -snN -nN -pmm 0 -pi 0.16 -pd 0 human_g1k_v37.fasta* |
| *mason illumina -N 100000 -i -o Set4_300.fasta -n 300 -f -snN -nN -pmm 0 -pi 0.16 -pd 0 human_g1k_v37.fasta* |
| **Set 5: (400,000 deletion-rich reads)** |
| *mason illumina -N 100000 -i -o Set5_64.fasta -n 64 -f -snN -nN -pmm 0 -pi 0 -pd 0.16 human_g1k_v37.fasta* |
| *mason illumina -N 100000 -i -o Set5_100.fasta -n 100 -f -snN -nN -pmm 0 -pi 0 -pd 0.16 human_g1k_v37.fasta* |
| *mason illumina -N 100000 -i -o Set5_150.fasta -n 150 -f -snN -nN -pmm 0 -pi 0 -pd 0.16 human_g1k_v37.fasta* |
| *mason illumina -N 100000 -i -o Set5_300.fasta -n 300 -f -snN -nN -pmm 0 -pi 0 -pd 0.16 human_g1k_v37.fasta* |
| **mrFAST: version 2.6.1.0** |
| *mrfast --search human_g1k_v37.fasta --seq ERR240727_1.filt.fastq -o ERR240727_1.map -e 20* |
| **SHD** |
| *countPassFilter <edit-distance threshold > <input pairs>* |
| **Adjacency Filter (part of mrFAST mapper)** |
| *mrfast --search human_g1k_v37.fasta --seq ERR240727_1.filt.fastq -o ERR240727_1.map -e <edit-distance threshold >* |

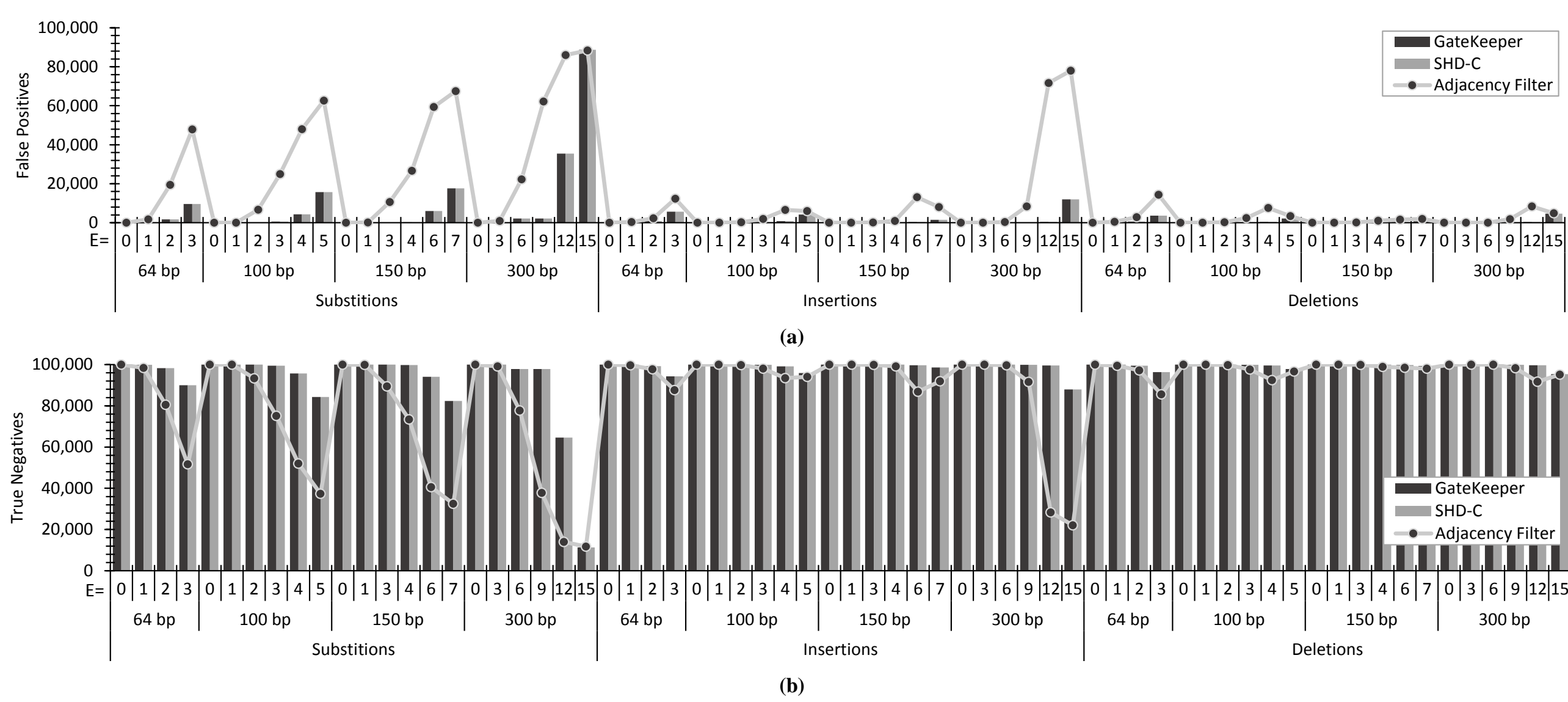

(a)

(b)

**Fig. 10: Breakdown of the number of (a) False positives and (b) True negatives of GateKeeper, SHD, and the Adjacency Filter across different edit distance thresholds (E) and read lengths using simulated Set3, Set4, and Set5.**

**Table 7: Total processing time breakdown of mrFAST with and without a pre-alignment step**

| | **Read Length = 100 bp, E = 5 edits** | **Read Length = 300 bp, E = 15 edits** |
|---|---|---|
| **mrFAST-2.1 (without pre-alignment)** | | |
| Candidate alignment locations | 4% | 8% |
| SIMD banded Levenshtein edit distance | 93% | 92% |
| SAM printing | 3% | 0% |
| Total run time (hours) | 24 hours | 1.02 hours |
| **mrFAST-2.6 (with the Adjacency Filter)** | | |
| Candidate alignment locations with Adjacency Filter | 32% | 71% |
| SIMD banded Levenshtein edit distance | 59% | 29% |
| SAM printing | 9% | 0% |
| Total run time (hours) | 7.3 hours | 0.12 hours |
| **mrFAST-2.1 (with GateKeeper)** | | |
| Candidate alignment locations | 49% | 96% |
| GateKeeper | 5% | 0.089% |
| SIMD banded Levenshtein edit distance | 17% | 4% |
| SAM printing | 29% | 0% |
| Total run time (hours) | 2.5 hours | 0.09 hours |

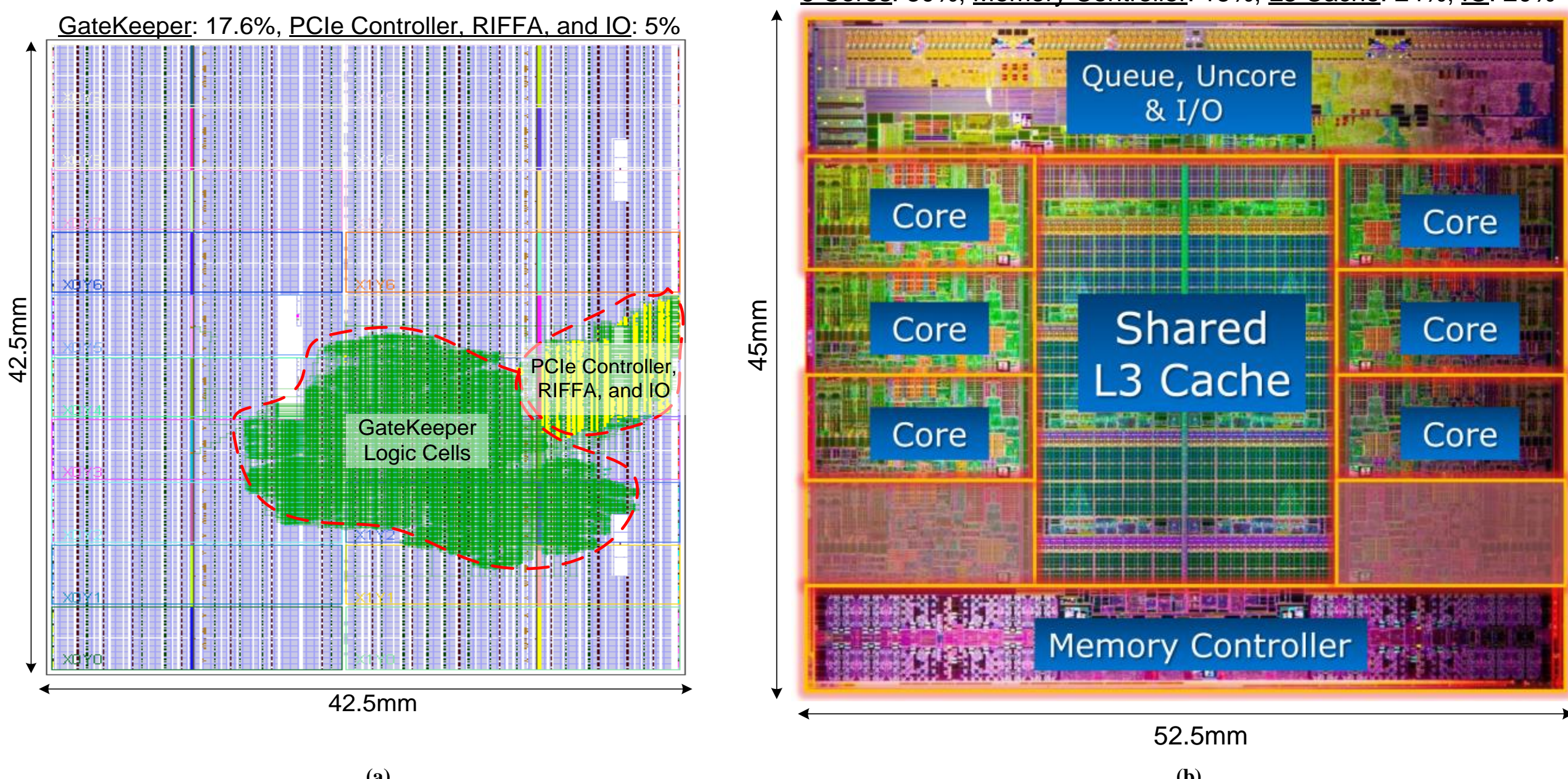

**Fig. 11: Chip layout and breakdown of the chip area for (a) GateKeeper (for a read length of 300 bp and E=15) built on VC709 FPGA and (b) Intel i7 CPU (Shimpi, 2011).**